\documentclass[11pt]{article}
\usepackage[margin=1in]{geometry}
\usepackage{graphicx}
\usepackage{amsmath}
\usepackage{hyperref}
\usepackage{longtable}
\usepackage{float}
\usepackage{caption}
\usepackage{booktabs}
\usepackage{authblk}

\title{\textbf{An 11,000-Study Open-Access Dataset of Longitudinal Magnetic Resonance Images of Brain Metastases}}
\usepackage{authblk}

\author[1,*]{Saahil Chadha}
\author[1,2,*]{David Weiss}
\author[3]{Anastasia Janas}
\author[4]{Divya Ramakrishnan}
\author[5,6]{Thomas Hager}
\author[1,7]{Klara Osenberg}
\author[1,7]{Klara Willms}
\author[5]{Joshua Zhu}
\author[8]{Veronica Chiang}
\author[9]{Spyridon Bakas}
\author[10,11]{Nazanin Maleki}
\author[5,6]{Durga V. Sritharan}
\author[12]{Sven Schoenherr}
\author[12]{Malte Westerhoff}
\author[1]{Matthew Zawalich}
\author[1]{Melissa Davis}
\author[1]{Ajay Malhotra}
\author[12]{Khaled Bousabarah}
\author[2]{Cornelius Deuschl}
\author[1,3]{MingDe Lin}
\author[5,6,13,14,15,$\dagger$]{Sanjay Aneja}
\author[10,11,$\dagger$]{Mariam S. Aboian}

\affil[1]{Department of Radiology and Biomedical Imaging, Yale School of Medicine, USA}
\affil[2]{Department of Diagnostic and Interventional Radiology and Neuroradiology, University Hospital Essen, Germany}
\affil[3]{Visage Imaging, Inc., San Diego, USA}
\affil[4]{Department of Radiology, Stanford University, USA}
\affil[5]{Department of Therapeutic Radiology, Yale School of Medicine, USA}
\affil[6]{Center for Outcomes Research and Evaluation, Yale New Haven Hospital, USA}
\affil[7]{Universitat Leipzig, Leipzig, Germany}
\affil[8]{Department of Neurosurgery, Yale School of Medicine, USA}
\affil[9]{Department of Pathology \& Laboratory Medicine, Indiana University School of Medicine, USA}
\affil[10]{Department of Radiology, Children’s Hospital of Philadelphia, USA}
\affil[11]{Department of Radiology, University of Pennsylvania Perelman School of Medicine, USA}
\affil[12]{Visage Imaging, GmbH, Berlin, Germany}
\affil[13]{Department of Biomedical Informatics \& Data Science, Yale School of Medicine, USA}
\affil[14]{Department of Biomedical Engineering, Yale University, USA}
\affil[15]{Yale Cancer Center, USA}
\affil[*]{These authors contributed equally}
\affil[$\dagger$]{These senior authors contributed equally}

\date{}

\begin{document}

\maketitle

Corresponding Author:
Mariam S. Aboian (\href{mailto:aboianm@chop.edu}{aboianm@chop.edu})

\section*{Abstract}
Brain metastases are a common complication of systemic cancer, affecting over 20\% of patients with primary malignancies. Longitudinal magnetic resonance imaging (MRI) is essential for diagnosing patients, tracking disease progression, assessing therapeutic response, and guiding treatment selection. However, the manual review of longitudinal imaging is time-intensive, especially for patients with multifocal disease. Artificial intelligence (AI) offers opportunities to streamline image evaluation, but developing robust AI models requires comprehensive training data representative of real-world imaging studies. Thus, there is an urgent necessity for a large dataset with heterogeneity in imaging protocols and disease presentation. To address this, we present an open-access dataset of 11,884 longitudinal brain MRI studies from 1,430 patients with clinically confirmed brain metastases, paired with clinical and image metadata. The provided dataset will facilitate the development of AI models to assist in the long-term management of patients with brain metastasis.

\section*{background and summary}
Brain metastasis is a common cancer complication, with more than 20\% of all patients with primary malignancies developing brain metastasis.1 These secondary brain tumors are associated with significant morbidity and mortality,1–3 and their management requires close radiologic monitoring and communication with the treating neuro-oncologists to assess treatment response and disease progression. 

Magnetic resonance imaging (MRI) plays a central role in the management of patients with brain metastases, guiding therapies such as stereotactic radiosurgery (SRS), whole-brain radiation therapy (WBRT), and surgical resection.4–10 Longitudinal MRI data is critical for tracking the behavior of brain metastases, guiding treatment selection, and assessing therapeutic response.11–13 This process often involves identifying subtle changes in tumor morphology, requiring careful comparison across numerous imaging studies. However, the manual review of longitudinal imaging presents a significant challenge for clinicians, given the time-intensive nature of simultaneously evaluating multiple lesions, sequences, and timepoints.14 

Artificial intelligence (AI) has emerged as a powerful resource for treatment planning and prognostication in neuro-oncology, offering the potential to accelerate the evaluation of longitudinal imaging, automate tumor segmentation, predict treatment response, and personalize therapeutic strategies.14–20  AI models have shown promise in enhancing medical imaging analysis; however, the development of generalizable and clinically robust models is limited by the scarcity of large, heterogeneous datasets.21,22 Optimal datasets must capture real-world variations in imaging protocols and disease presentation to ensure broad applicability and performance across disparate patient populations. 

Thus, there is a critical need for a longitudinal dataset that is heterogenous with respect to imaging phenotypes, scanners, and acquisition parameters. We present such a dataset of 11,884 brain MRI studies from 1,430 patients with radiologically or pathologically confirmed brain metastases. This dataset spans nearly two decades of imaging, capturing pre- and post-treatment timepoints with four essential MRI sequences: T1-weighted pre-contrast (T1W), T1-weighted post-contrast (T1CE), T2-weighted (T2), and fluid-attenuated inversion recovery (FLAIR). Accompanied by detailed clinical metadata, this resource offers a unique opportunity to study disease progression, treatment response, and imaging biomarker development.

To the authors’ knowledge, this dataset represents the largest public release of MRI from patients with brain metastases. We anticipate that this open-access dataset will facilitate a wide range of research applications, from traditional radiologic studies to advanced machine learning techniques, ultimately contributing to improved patient outcomes and a deeper understanding of brain metastases. This release is aligned with the AI for Response Assessment in Neuro Oncology recommendations for good clinical practice in the development and validation of AI tools in neuro-oncology, emphasizing transparency, reproducibility, and clinical relevance.23,24 By leveraging both imaging and clinical data, this dataset serves as a valuable resource for the oncology, neuroradiology, and data science communities.

\section*{Methods} 
\subsection*{Subject Characteristics}

The electronic medical record (EMR) at Yale New Haven Hospital was queried for MRI studies conducted between 2004 and 2023 that assessed the presence of brain metastasis. This automated search identified a total of 46,364 studies from 7,111 patients that potentially exhibited intracranial metastatic disease. A manual review of the electronic health record (EHR) system excluded patients without radiologic or pathologic evidence of brain metastasis from the initial dataset. Additionally, only studies with axial T1W, T1CE, T2, or FLAIR MRI sequences were included. For patients who received brain metastasis-directed treatment (e.g. SRS, WBRT, or surgical resection), only pre-treatment scans taken within 30 days before treatment initiation were included, along with all available follow-up scans, for longitudinal assessment of brain metastasis development and treatment response. The final dataset includes a total of 11,884 MRI studies from 1,430 patients with clinical evidence of brain metastasis. A summary of the demographic and clinical characteristics is provided in Table 1, with additional details available in the accompanying Excel file. 
This dataset is available on The Cancer Imaging Archive (TCIA), offering public access to MRI sequences and clinical metadata. The composition and curation process of the dataset are illustrated in Figure 1.

\subsection*{Patient and Study Data}

Baseline information for each patient was obtained from the EMR at every study timepoint. The collected data includes the patient’s age at the time of the imaging study, sex, and study date. This data was retrieved as of December 2023.

\subsection*{MRI Characteristics }

The majority of images were acquired using 1.5 T or 3 T MRI scanners from Siemens Healthineers or General Electric HealthCare. Imaging data and metadata for the included studies were extracted via the application programming interface (API) from Visage (Visage 7, Visage Imaging, Inc., San Diego, CA). DICOM metadata facilitated the extraction of key image acquisition characteristics, including study site, scanner vendor, scanner model, field strength, two-dimensional (2D)/three-dimensional (3D) acquisition type, sequence name, slice thickness, slice spacing, repetition time (TR), echo time (TE), and inversion time (TI). An overview of imaging acquisition parameters is provided in Table 2, with a detailed list for each scan in the accompanying Excel file.

\subsection*{MRI Sequence Selection and Standardization} 

T1W, T1CE, T2, and FLAIR MRI sequences were selected for inclusion, as they offer complementary imaging features essential for diagnosis and longitudinal monitoring of brain metastases. MRI sequence naming was standardized to account for variations in DICOM metadata between scanners, radiology technicians, and sites. Manual review of studies was implemented to develop a rules-based image classifier and validator. Images were filtered based on orientation, sequence acquisition, contrast-relation, and spin echo variation, ensuring that only relevant sequences were included. Naming was further standardized by removing redundancy in sequence identifiers across the dataset. This systematic approach ensured the accurate inclusion and consistent labeling of MRI sequences, optimizing the dataset for longitudinal analysis.

\subsection*{Anonymization} 

All MRI studies were copied and de-identified on the clinical Picture Archiving and Communication System (PACS) archive. This process removed DICOM metadata with protected health information, retaining anonymous identifiers for each imaging study and patient. The de-identified studies were then sent to a secure research PACS (AI Accelerator [AIA] Visage Imaging, Inc., San Diego, CA). Included studies were exported to a secure external drive as NIfTI files using the Visage API. After sequence selection and standardization, HD-BET was used to extract brain parenchyma from each image file, removing identifiable facial features.

\subsection*{Ethical Approval }

This retrospective study was approved by the Institutional Review Board of Yale University on 10/01/2020, protocol 2000029055. 
\section*{Data Records
}
The data records have been submitted to TCIA collections. MR images are provided alongside an Excel file with separate spreadsheets for clinical data and radiologic image acquisition parameters. For each brain metastasis study, up to four files are provided with T1W, T1CE, T2W, and/or FLAIR sequence images. All sequences were exported from the AIA in NIfTI format and brain-extracted. All file names include an anonymous patient identifier, anonymous study date-time, and sequence type, formatted as caseID\_date-time\_sequence.nii.gz. 
\section*{Technical Validation}
The final dataset included only patients with T1W, T1CE, T2W, and FLAIR images. Several deep learning-based brain extraction techniques have been developed.25,26 In this study, brain extraction was performed using HD-BET,25 which has demonstrated high performance across various MRI sequences. The quality of the brain extraction was manually verified on a random sample of 200 studies, ensuring that no skull residues or incorrect brain masks were present.

Descriptive statistics were also calculated for key acquisition parameters, such as slice thickness, TR, and TE, and showed consistency in distribution across sites and scanners. Figure 2 compares image acquisition parameter distribution by sequence and site, and Figure 3 provides an analysis of the longitudinal robustness of the dataset.

\section*{Usage Note}
All files can be downloaded from TCIA public collections (https://www.cancerimagingarchive.net/ https://doi.org/10.7937/3YAT-E768). Image files can be opened on segmentation platforms that support NIfTI format. 
\section*{Code Availability }
Code used for MRI sequence standardization and brain extraction is available upon request.
\section*{Acknowledgments}Research reported in this publication was supported by the Richard K. Gershon Endowed Medical Student Research Fellowship and Yale School of Medicine Fellowship for Medical Student Research. The content is solely the responsibility of the authors and does not necessarily represent the official views of the Richard K. Gershon Endowed Medical Student Research Fellowship and Yale School of Medicine Fellowship for Medical Student Research. 
\section*{Author contributions }
M.S.A contributed to project conception. M.S.A, S.A., and N.M. were involved in project oversight. D.W. and A.J. contributed data curation and quality control. S.C. and T.H. conducted data export and processing. S.C. contributed to data analysis. S.C. and D.W. drafted the original manuscript. All authors contributed to and approved the final manuscript.
\section*{Competing interests }
M.L. and K.B. are employees of Visage Imaging, and M.L is a stockholder of Visage Imaging. All other authors declare no competing interests. 
\begin{figure}[H]
\centering
\includegraphics[width=\textwidth]{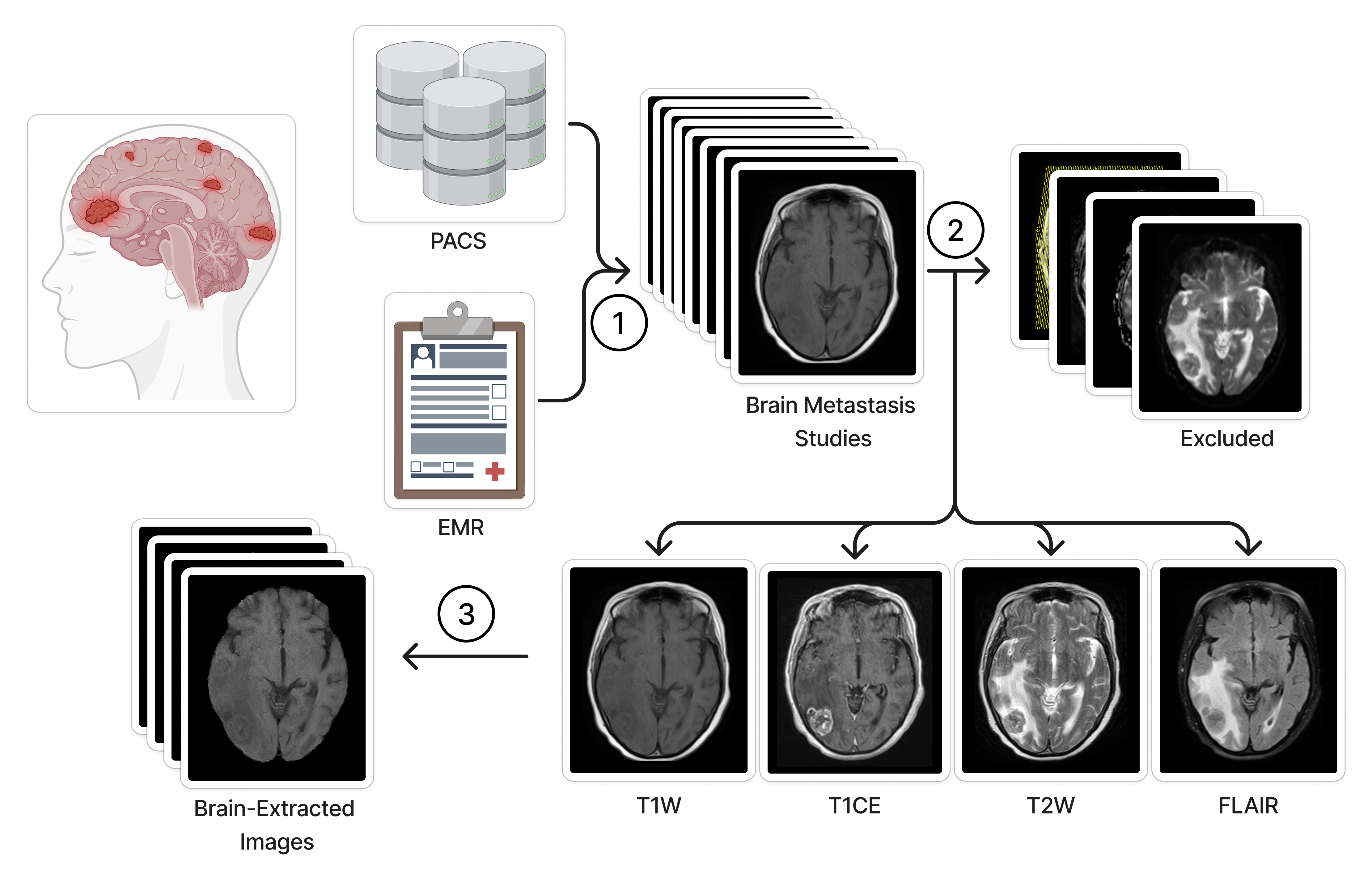}
\caption{Data Processing Workflow. This figure outlines the processes of retrospective MRI study
curation. In step (1) all MRI studies evaluating for the presence or absence of brain metastasis
were queried and retrieved from the PACS archive. The EHR was then manually reviewed and
only the studies with evidence of brain metastasis were retained. In step (2), the remaining studies
were processed via a sequence classifier that utilized study metadata. Only T1W, T1CE, T2W, and
FLAIR sequences were included, and other acquired sequences were discarded. In step (3) HD-BET
was used to remove all extra-parenchymal tissue and protect patient privacy. The resulting dataset
was uploaded to TCIA.}
\label{fig:workflow}
\end{figure}

\begin{figure}[H]
\centering
\includegraphics[width=\textwidth]{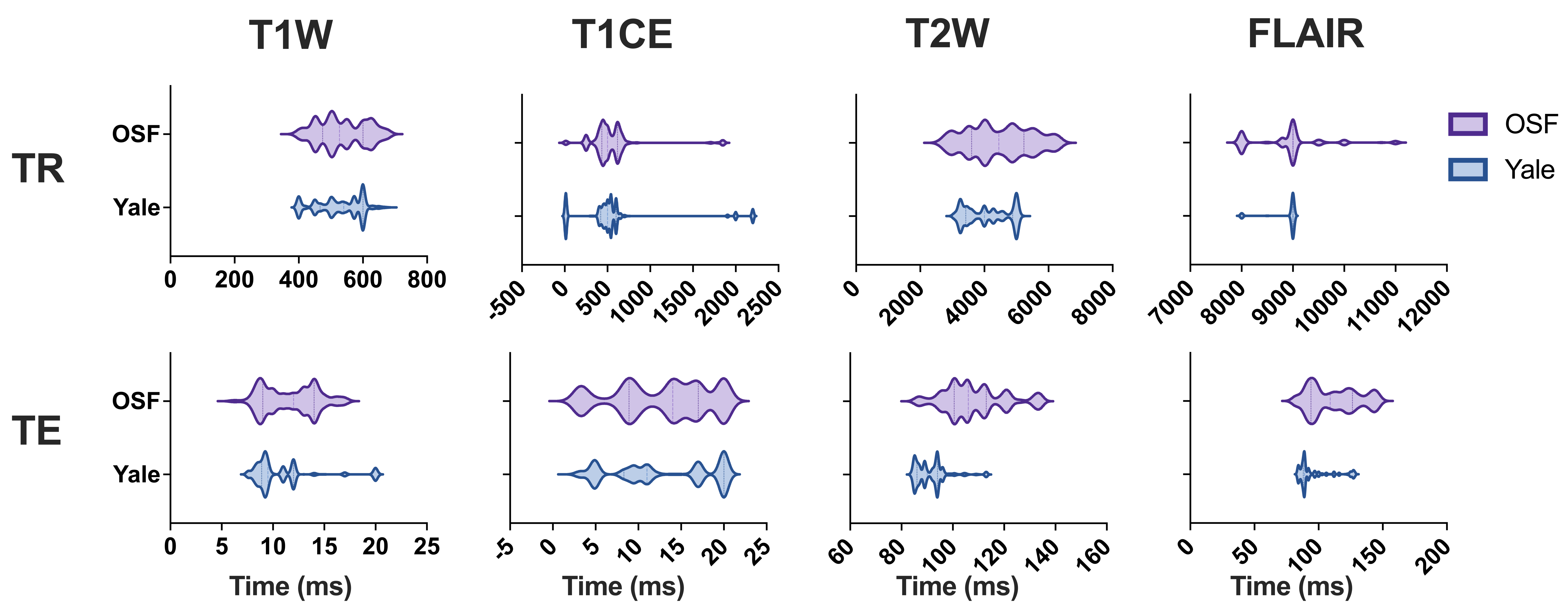}
\caption{Comparison of Image Acquisition Parameters by Sequence and Scanner Site. The
top row displays the distribution or repetition time (TR) at scans acquired at Yale and outside
facilities (OSF), for T1W, T1CE, T2W, and FLAIR sequences (from left to right). The bottom
row displays the distributions of echo time (TE).}
\label{fig:params}
\end{figure}

\begin{figure}[H]
\centering
\includegraphics[width=\textwidth]{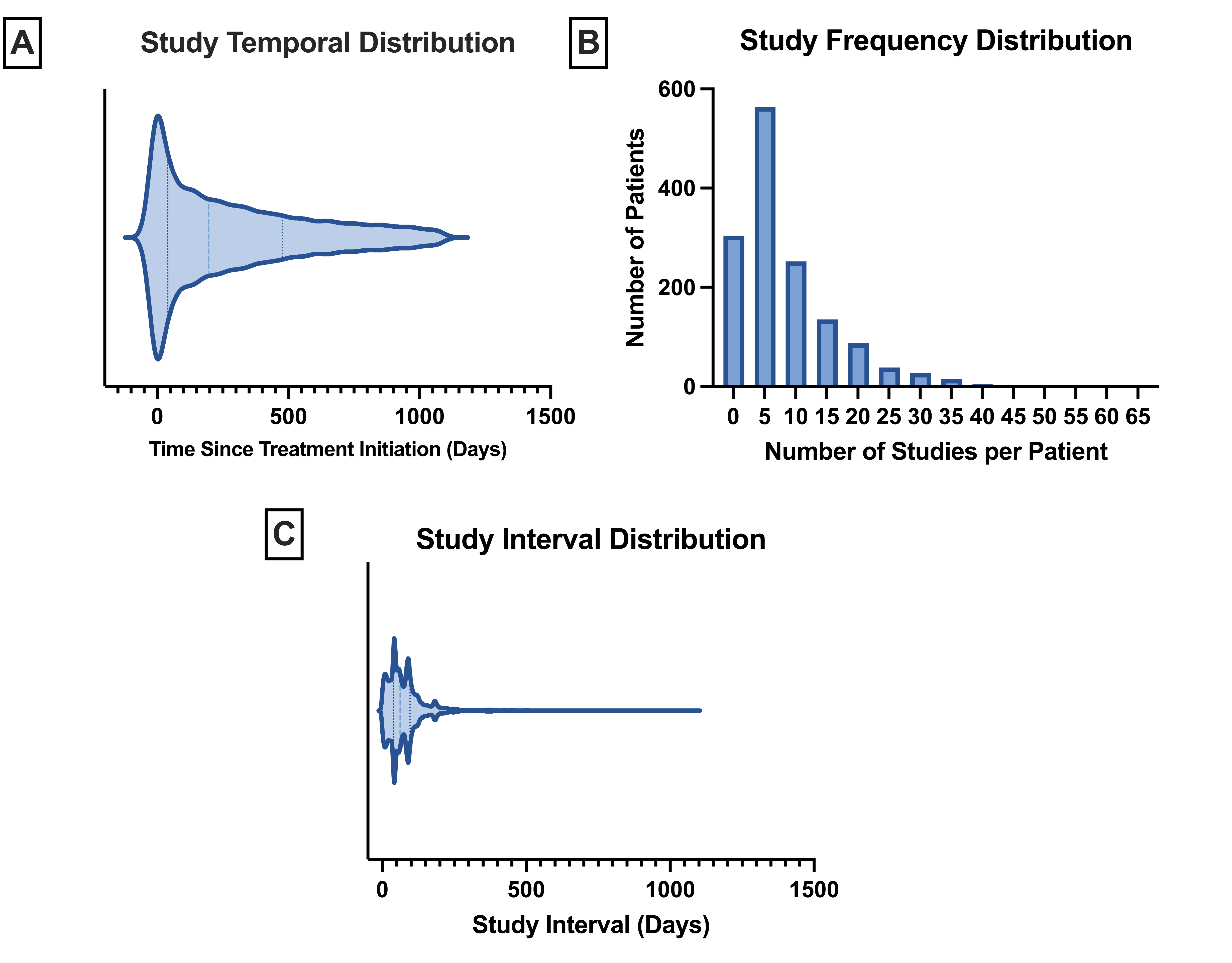}
\caption{Longitudinal Study Analysis. (A) is a distribution of studies in the dataset by
time since treatment initiation. (B) provides a histogram of the number patients who have varying
number of study time-points included in the dataset. (C) shows a distribution of follow-up intervals
between each patient’s studies.}
\label{fig:longitudinal}
\end{figure}

\begin{table}[H]
\centering
\begin{minipage}[t]{0.45\textwidth}
\centering
  \caption{Baseline Demographic Information }  
\vspace{0.5em} 
\small 

\begin{tabular}{ll}
\toprule
\ & \textbf{Summary}\\\midrule
Number of patients & 1,430 \\
Sex & \\
\quad Male (\%) & 611 (43\%) \\
\quad Female (\%) & 819 (57\%) \\
Age at earliest study (median; IQR) & 64 (56–71) \\
Number of studies & 11,892 \\
Treatment status & \\
\quad Pre-treatment (\%) & 1,633 (15\%) \\
\quad Follow-up (\%) & 9,455 (85\%) \\
Obtained at OSF (\%) & 803 (7\%) \\
\bottomrule
\end{tabular}
\end{minipage}
\hfill
\begin{minipage}[t]{0.45\textwidth}
\centering
\caption{Scanner Details}
\vspace{0.5em} 
\small 
\begin{tabular}{ll}
\toprule
\ & \textbf{Summary} \\\midrule
Model & \\
\quad Siemens Avanto (\%) & 2,637 (22\%) \\
\quad Siemens Espree (\%) & 2,152 (18\%) \\
\quad Siemens Verio (\%) & 3,352 (28\%) \\
\quad Other Siemens (\%) & 2,110 (18\%) \\
\quad GE Signa Excite (\%) & 942 (8\%) \\
\quad GE Signa HDxt (\%) & 200 (2\%) \\
\quad Other GE (\%) & 351 (3\%) \\
\quad Other scanner models (\%) & 143 (1\%) \\
Magnet Strength & \\
\quad 3 T (\%) & 4,931 (41\%) \\
\quad 1.5 T (\%) & 6,839 (58\%) \\
\quad Other (\%) & 66 (0.6\%) \\
3D Acquisition (\%) & 2,381 (20\%) \\
\bottomrule
\end{tabular}
\end{minipage}
\end{table}
\begin{table}[H]
\centering
\caption{Image Acquisition Parameters. SD indicates standard deviation.}
\begin{tabular}{lcccc}
\toprule
 & \textbf{T1W} & \textbf{T1CE} & \textbf{T2W} & \textbf{FLAIR} \\\midrule
Number of studies with sequence available (\%) & 8,392 (71\%) & 9,003 (76\%) & 7,354 (62\%) & 9,090 (76\%) \\
TR (ms) (mean ± SD) & 586 ± 314 & 591 ± 547 & 4,130 ± 902 & 8,841 ± 587 \\
TE (ms) (mean ± SD) & 10.8 ± 4.1 & 12.6 ± 6.6 & 94.9 ± 22.6 & 100.6 ± 31.5 \\
TI (ms) (mean ± SD) & 259 ± 440 & 437 ± 463 & 0 ± 0 & 2,432 ± 170 \\
\bottomrule
\end{tabular}
\end{table}

References 

1.	Achrol, A. S. et al. Brain metastases. Nat Rev Dis Primers 5, 5 (2019). 

2.	Sacks, P. \& Rahman, M. Epidemiology of Brain Metastases. Neurosurgery Clinics of North America 31, 481–488 (2020). 

3.	Lamba, N., Wen, P. Y. \& Aizer, A. A. Epidemiology of brain metastases and leptomeningeal disease. Neuro Oncol 23, 1447–1456 (2021). 

4.	Brenner, A. W. \& Patel, A. J. Review of Current Principles of the Diagnosis and Management of Brain Metastases. Front Oncol 12, 857622 (2022). 

5.	Lin, X. \& DeAngelis, L. M. Treatment of Brain Metastases. J Clin Oncol 33, 3475–3484 (2015). 

6.	Vogelbaum, M. A. et al. Treatment for Brain Metastases: ASCO-SNO-ASTRO Guideline. J Clin Oncol 40, 492–516 (2022). 

7.	Le Rhun, E. et al. EANO-ESMO Clinical Practice Guidelines for diagnosis, treatment and follow-up of patients with brain metastasis from solid tumours. Ann Oncol 32, 1332–1347 (2021). 

8.	Patil, C. G. et al. Whole brain radiation therapy (WBRT) alone versus WBRT and radiosurgery for the treatment of brain metastases. Cochrane Database Syst Rev 9, CD006121 (2017). 

9.	Kraft, J., Zindler, J., Minniti, G., Guckenberger, M. \& Andratschke, N. Stereotactic Radiosurgery for Multiple Brain Metastases. Curr Treat Options Neurol 21, 6 (2019). 

10.	Aldawsari, A. M. et al. The role and potential of using quantitative MRI biomarkers for imaging guidance in brain cancer radiotherapy treatment planning: A systematic review. Phys Imaging Radiat Oncol 27, 100476 (2023). 

11.	Kang, T. W. et al. Morphological and functional MRI, MRS, perfusion and diffusion changes after radiosurgery of brain metastasis. Eur J Radiol 72, 370–380 (2009). 

12.	Friedman, D. P., Morales, R. E. \& Goldman, H. W. MR imaging findings after stereotactic radiosurgery using the gamma knife. AJR Am J Roentgenol 176, 1589–1595 (2001). 

13.	Lunsford, L. D., Kondziolka, D., Maitz, A. \& Flickinger, J. C. Black holes, white dwarfs and supernovas: imaging after radiosurgery. Stereotact Funct Neurosurg 70 Suppl 1, 2–10 (1998). 

14.	Cassinelli Petersen, G. et al. Real-time PACS-integrated longitudinal brain metastasis tracking tool provides comprehensive assessment of treatment response to radiosurgery. Neurooncol Adv 4, vdac116 (2022). 

15.	Aneja, S. \& Omuro, A. Imaging biomarkers for brain metastases: more than meets the eye. Neuro Oncol 21, 1493–1494 (2019). 

16.	Aneja S \& Krumholz HK. Predicting Radiation Side Effects Using Deep Learning. Radiology: Artificial Intelligence (2018). 

17.	Aboian, M. et al. Development of a workflow efficient PACS based automated brain tumor segmentation and radiomic feature extraction for clinical implementation (N2.003). Neurology 98, (2022). 

18.	Aboian, M. et al. Clinical implementation of artificial intelligence in neuroradiology with development of a novel workflow-efficient picture archiving and communication system-based automated brain tumor segmentation and radiomic feature extraction. Frontiers in Neuroscience 16, (2022). 

19.	Rudie, J. D. et al. Three-dimensional U-Net Convolutional Neural Network for Detection and Segmentation of Intracranial Metastases. Radiol Artif Intell 3, e200204 (2021). 

20.	Xue, J. et al. Deep learning-based detection and segmentation-assisted management of brain metastases. Neuro Oncol 22, 505–514 (2020). 

21.	Aneja, S., Chang, E. \& Omuro, A. Applications of artificial intelligence in neuro-oncology. Current Opinion in Neurology 32, 850 (2019). 

22.	Rudie, J. D., Rauschecker, A. M., Bryan, R. N., Davatzikos, C. \& Mohan, S. Emerging Applications of Artificial Intelligence in                     Neuro-Oncology. Radiology 290, 607–618 (2019). 

23.	Isensee, F. et al. Automated brain extraction of multisequence MRI using artificial neural networks. Hum Brain Mapp 40, 4952–4964 (2019).

\typeout{get arXiv to do 4 passes: Label(s) may have changed. Rerun}

\end{document}